\documentclass[reqno]{amsart}
\usepackage{mathrsfs}
\usepackage{cite}
\newtheorem{remark}{Remark}

\begin{document}

\title[Matrix generalizations of integrable systems]{Matrix generalizations of integrable systems with Lax integro-differential representations}
\author{O. Chvartatskyi$^1$, Yu. Sydorenko$^1$}%

\address{$^1$Department of Mechanics and Mathematics, Ivan Franko National University of Lviv, 1, Universytetska str., Lviv, 79000, Ukraine}

\email{alex.chvartatskyy@gmail.com, y$\_$sydorenko@franko.lviv.ua}

\begin{abstract}
%%% some changes

We present (2+1)-dimensional generalizations of the
k-constrained Kadomtsev-Petviashvili (k-cKP) hierarchy
and corresponding matrix Lax representations that consist of two
integro-differential operators. Additional reductions
imposed on the Lax pairs lead to matrix generalizations of
Davey-Stewartson systems (DS-I,DS-II,DS-III) and (2+1)-dimensional extensions
of the modified Korteweg-de Vries and the Nizhnik equation. We also
present an integro-differential Lax pair for a matrix version of a
(2+1)-dimensional extension of the Chen-Lee-Liu equation.
\end{abstract}
\maketitle
\section{Introduction}
In the modern theory of nonlinear integrable systems, algebraic
methods play an important role (see the survey in \cite{LDA}). Among
them are the Zakharov-Shabat dressing method
\cite{Zakharov,Zakh-Manak,solitons}, Marchenko's method
\cite{March}, and the approach based on Darboux-Crum-Matveev
transformations \cite{Matveev79,Matveev}. Algebraic methods allow to
omit analytical difficulties that arise in the investigation of
corresponding direct and inverse scattering problems for nonlinear
equations. A significant contribution to such methods has also been
made by the Kioto group \cite{DJKM1,DJKM2,SS3,MM4,Ohta}. In
particular, they investigated scalar and matrix hierarchies for
nonlinear integrable systems of Kadomtsev-Petviashvili type (KP
hierarchy).

The KP hierarchy %(see e.g. \cite{LDA,SS3,Ohta})
%%% perhaps you insert some more references
is of fundamental importance in the theory of integrable systems and shows up
in various ways in mathematical physics.
Several extensions and generalizations of it have been obtained.
For example, the multi-component KP
%%% (mcKP)          NOT needed !
hierarchy contains several physically relevant nonlinear integrable systems,
including the Davey-Stewartson equation, the two-dimensional Toda lattice and
the three-wave resonant interaction system. There are several equivalent formulations
of this hierarchy: matrix pseudo-differential operator (Sato) formulation,
$\tau$-function approach via matrix Hirota bilinear identities,
multi-component free fermion formulation.
%%% You may insert some references here, e.g. the book
%%%  Solitons: Differential Equations, Symmetries and Infinite Dimensional Algebras by Miwa, Jimbo, Date
%%% and Kupershmidt's book "KP or mKP"
Another kind of generalization is the so-called ``KP equation with
self-consistent sources'' (KPSCS), discovered by Melnikov
\cite{M1,Mf,M4,M2,M3}. In \cite{SS,KSS,Chenga1,CY,Chenga2},
k-symmetry constraints of the KP hierarchy were investigated, which
have connections with KPSCS.
%%%
The resulting k-constrained KP (k-cKP) hierarchy contains physically
relevant systems like the nonlinear Schr\"odinger equation, the
Yajima-Oikawa system, a generalization of the Boussinesq equation,
and the Melnikov system. Multi-component generalizations of the
k-cKP hierarchy were considered in \cite{SSq}. In papers
\cite{Oevel93,Oevel96,Aratyn97} the differential type of the gauge
transformation operator was applied to the constrained KP hierarchy
at first. A modified k-constrained KP (k-cmKP) hierarchy was
proposed in \cite{CY,KSO,OC}. It contains, for example, the vector
Chen-Lee-Liu and the modified KdV (mKdV) equation. Multi-component
versions of the Kundu-Eckhaus and Gerdjikov-Ivanov equations were
also obtained in \cite{KSO}, via gauge transformations of the k-cKP,
respectively the k-cmKP hierarchy.

Moreover, in \cite{MSS,6SSS}, (2+1)-dimensional extensions of the
k-cKP hierarchy were introduced. In \cite{BS1,PHD}, exact solutions
for some representatives of the (2+1)-dimensional k-cKP hierarchy
were obtained by dressing binary transformations. (2+1)-dimensionalç
extensions of k-cKP and k-cmKP and their dressings with the help of
differential transformations were investigated in \cite{LZL1,LZL2}.
The (2+1)-dim\-en\-si\-o\-n\-al k-cKP hierarchy in particular
contains the DS-III system and a (2+1)-dimensional extension of the
mKdV equation. A corresponding Lax representation of the
(2+1)-dim\-ensio\-nal k-cKP hierarchy consists of one differential
and one integro-differential operator. Our aim was to generalize Lax
representations of the (2+1)-dim\-ensi\-on\-al k-cKP hierarchy to
the case of two integro-differential operators, in order to obtain
Lax representations for matrix generalizations of Davey-Stewartson
systems DS-I, DS-II,
%%% The relation between "two integro-differential operators" and "matrix" is not clear
DS-III, and their higher order counterparts.
%%% symmetries.
We also present a Lax representation, with two integro-differential operators, for
a (2+1)-dimensional generalization of the Chen-Lee-Liu equation, which has been
obtained in \cite{OC}.

 This work is organized as follows. In Section 2 we
 present a short survey of results on constraints for KP hierarchies
 and their (2+1)-dimensional generalizations. In Sections 3 and 4 we consider
 integro-differential Lax representations that generalize
 corresponding representations for the (2+1)-dimensional k-cKP hierarchy. As a
 result of additional reductions, we obtain matrix generalizations of Davey-Stewartson
 and (2+1)-dimensional mKdV equations that generalize corresponding
 systems in the (2+1)-dimensional k-cKP hierarchy. In Section 5, by application
 of a gauge transformation, we obtain a Lax representation for a (2+1)-dimensional matrix
extension of the Chen-Lee-Liu equation.
In the final section,
%%%Conclusions
we discuss the obtained results and mention problems for further investigations.

\section{k-constrained KP hierarchy and its extensions}

To make this paper somewhat self-contained, we briefly introduce the
KP hierarchy \cite{LDA}, its k-symmetry constraints (k-cKP
hierarchy), and the extension of the k-cKP hierarchy to the
(2+1)-dimensional case \cite{MSS,6SSS}. A Lax representation of the
KP hierarchy is given by
\begin{equation}
L_{t_n}=[B_n,L],\,    \qquad n\geq1,
\end{equation}
where $L=D+U_1D^{-1}+U_2D^{-2}+\ldots$ is a scalar
pseudodifferential operator, $t_1:=x$, $D:=\frac{\partial}{\partial
x}$, and $B_n:= (L^n)_+
 := (L^n)_{\geq0}=D^n+\sum_{i=0}^{n-2}u_iD^i$ is
the differential operator part of $L^n$. The consistency condition (zero-curvature equations),
arising from the commutativity of the flows,
%%% $\partial_{t_n}$ and $\partial_{t_k}$
%%% gives rise to the zero-curvature equations of the KP hierarchy:
are
\begin{equation}
B_{n,t_k}-B_{k,t_n}+[B_n,B_k]=0.
\end{equation}
%%%It is well-known

Let $B^{\tau}_n$ denote the formal transpose of $B_n$, i.e.
$B^{\tau}_n:=(-1)^nD^n+\sum_{i=0}^{n-2}(-1)^iD^iu^{\top}_i$, where
$^{\top}$ denotes the matrix transpose. We use curly brackets to
denote the action of an operator on a function whereas, for example,
$B_n \, q$ means composition of the operator $B_n$ and the operator
of multiplication by the function $q$. The k-cKP hierarchy
\cite{SS,KSS,Chenga1,CY,Chenga2} is given by
\begin{equation}\label{eq1}
\begin{array}{c}
  L_{t_n}=[B_n,L], \quad
  q_{t_n}=B_n\{q\}, \quad
  -r_{t_n}=B^{\tau}_n\{r\}, \quad
\end{array}n=2,3,\ldots,
\end{equation}
 with the k-symmetry reduction
%%% for the Lax operator $L$:
\begin{equation}
    L_k:=L^k=B_k+{q}D^{-1}{r}.
\end{equation}
It admits the Lax representation (here $k\in{\mathbb{N}}$ is fixed)
\begin{equation}
  [B_k+qD^{-1}r,\partial_{t_n}-B_n]=0.
\end{equation}

Below we will also use the formal adjoint
$B^*_n:=\bar{B}^{\tau}_n=(-1)^nD^n+\sum_{i=0}^{n-2}(-1)^iD^i{u}^*_i$
of $B_n$, where $^\ast$ denotes Hermitian conjugation (complex
conjugation and transpose).

%where $q$ and $r$ are scalar functions satisfying the system:
%\begin{equation}
%\left\{\begin{array}{c}q_{t_k}=B_k\{q\},\\
%-r_{t_k}=B^{\tau}_k\{r\}.\end{array}\right.
%\end{equation}
%where $q$ and $r$ are scalar functions, is compatible with the KP
%hierarchy and reduces the KP hierarchy to the k-constrained KP
%hierarchy.

 In \cite{SSq}, multi-component (vector) generalizations of the
 k-cKP hierarchy were introduced,
\begin{equation}
\begin{array}{c}L^k=B_k+\sum_{i=1}^l\sum_{j=1}^lq_im_{ij}D^{-1}r_j=B_k+{\bf q}{\mathcal M}_0D^{-1}{\bf
r}^{\top},
\\{\bf q}_{t_n}=B_n\{{\bf q}\},\,\,{\bf r}_{t_n}=-B^{\tau}_n\{{\bf r}\} ,
\end{array}
\end{equation}
where ${\bf q}=(q_1,\ldots,q_l)$ and ${\bf r}=(r_1,\ldots,r_l)$ are
vector functions, ${\mathcal M}_0=(m_{ij})_{i,j=1}^l$ is a constant
$l\times l$ matrix. A corresponding Lax representation is given by
\begin{equation}\label{hier}
  [L_k,M_n]=0, \quad
  L_k=B_k+{\bf q}{\mathcal M}_0D^{-1}{\bf r}^{\top}, \quad
  M_n=\partial_{t_n}-B_n.
\end{equation}
 For $k=1$, this is a multi-component generalization of the AKNS
hierarchy. For $k=2$ and $k=3$, one obtains vector generalizations
of the Yajima-Oikawa and Melnikov \cite{Mf,M4} hierarchies,
respectively.

In \cite{CY,KSO,OC}, a k-constrained modified KP (k-cmKP) hierarchy
was introduced and investigated. Its Lax representation has the form
\begin{equation}
[\tilde{B}_k+{\bf q}{\mathcal M}_0D^{-1}{\bf
r}^{\top}D,\partial_{t_n}-\tilde{B}_n]=0,
\end{equation}
%%%% NOTE: I replaced w_j by w_{kj}
where $\tilde{B}_k=D^k+\sum_{j=1}^{k-1}w_{kj} D^j$.
%%% ($\tilde{B}_n$ has analogous form).
For $k=1,2,3$, this leads to vector generalizations of the Chen-Lee-Liu, the modified multi-component
Yajima-Oikawa and Melnikov hierarchies.

An essential extension of the k-cKP hierarchy is its
(2+1)-dimensional generalization \cite{MSS,6SSS}, given by
\begin{eqnarray}\label{spa1}
&& [L_k,M_n]=0, \quad
 L_k=\beta_k\partial_{\tau_k}-B_k-{\bf q}{\mathcal M}_0D^{-1}{\bf r}^{\top}, \quad
 M_n=\alpha_n\partial_{t_n}-{A}_n, \nonumber \\
&& B_k=D^k+\sum_{j=0}^{k-2}u_jD^j, \quad
 {A}_n=D^n+\sum_{i=0}^{n-2}v_iD^i, \quad \nonumber \\
&& u_j=u_j(x,\tau_k,t_n),\,\,v_i=v_i(x,\tau_k,t_n), \quad \alpha_n,\beta_k\in{\mathbb{C}},
\end{eqnarray}
where $u_j$ and $v_i$ are scalar functions, ${\bf q}$ and ${\bf r}$
are $l$-component vector-functions. An equivalent system is
$$ %\left\{
\begin{array}{lcl}
\alpha_{n} { {B_{k,t_n}}}=\beta_kA_{n,\tau_k}+[A_{n}, B_{k}] +
\left( [A_{n}, {\bf q} {\mathcal M}_0{D}^{-1} {\bf r}^{\top}]
\right)_{\geq0},
 \\
\alpha_{n}  {\bf q}_{t_n} =A_{n}\{{\bf q}\},\quad
 \alpha_{n} {{\bf r}}_{t_n}=- A_{n}^{\tau}\{{\bf r}\}.
\end{array}% \right.
$$
We list some members of this (2+1)-dimensional generalization of the k-cKP hierarchy:
\renewcommand{\theenumi}{\arabic{enumi}}
\begin{enumerate}
\item
 $k=1$, $n=2$. Then (\ref{spa1}) has the form
\begin{equation}\label{L}
 \left[ L_1, M_2\right] =0, \,
 L_1=\beta_{1} \partial_{\tau_{1}} - {D} -{\bf q}{\mathcal M}_0 {D}^{-1} {\bf r}^{\top},
 \,\,
 M_2=\alpha_{2} \partial_{t_{2}} - {D}^{2} - v_0,
\end{equation}
and it is equivalent to the following system,
\begin{equation}\label{Dsw}
\begin{array}{lcl}
 \alpha_{2} {\bf q}_{t_2}={\bf q}_{xx} + v_0{\bf q}, \quad
 \alpha_{2}  {\bf r}_{t_2}=-{\bf r}_{xx} - v_0{\bf r}, \\
\beta_{1} {v}_{0,\tau_1}={v}_{0,x} - 2 ({\bf q}{\mathcal M}_0{\bf
r}^{\top})_{x}.
\end{array}
\end{equation}
 After the reduction $\beta_1\in{\mathbb{R}}$, $\alpha_2\in i{\mathbb{R}}$,
${\bf r}=\bar{\bf q}$, ${\mathcal M}_0={\mathcal M}^*_0$,
$v_0=\bar{v}_0$, the operators $L_1$ and $M_2$ in (\ref{L}) are
skew-Hermitian and Hermitian, respectively, and (\ref{Dsw}) becomes
the DS-III system
\begin{equation}\label{DeS}
\begin{array}{lcl}
\alpha_{2} {\bf q}_{t_2}={\bf q}_{xx} + v_0{\bf q},\quad \beta_{1}
v_{0,\tau_1}=v_{0,x} - 2 ({\bf q}{\mathcal M}_0{\bf q}^*)_{x}.
\end{array}
\end{equation}

\item
$k=1$, $n=3$. Now (\ref{spa1}) becomes
\begin{equation}\label{L1M3e}
\begin{array}{l}
\left[ L_1, M_3\right] =0, \quad L_1=\beta_{1} \partial_{\tau_{1}} -
{D} - {\bf q}{\mathcal M}_0 {D}^{-1} {\bf r}^{\top},\,\,\\
M_3=\alpha_{3}\partial_{t_{3}} -{D}^3- v_{1} {D} - v_{0}.
\end{array}
\end{equation}
After the additional reduction $\alpha_3,\beta_1\in{\mathbb{R}}$,
${\mathcal M}_0={\mathcal M}^*_0$, $v_1=\bar{v}_1$,
$\bar{v}_0+v_0=v_{1x}$, the operators $L_1$, $M_3$ in (\ref{L1M3e})
are skew-Hermitian, and the Lax equation (\ref{L1M3e}) is equivalent
to the following (2+1)-dimensional generalization of the mKdV
system,
\begin{equation}\label{mKDV}
%\left\{
\begin{array}{lc}
\alpha_{3}{\bf q}_{t_3}  =  {\bf q}_{xxx} + v_{1}{\bf q}_{x} +
v_{0}{\bf q},\\ \beta_{1}  {v_{0,\tau_1}}
 =  {v_{0,x}} - 3 ({\bf q}_{x}{\mathcal M}_0{\bf q}^*)_{x}, \,\,
\beta_{1}  {v_{1,\tau_1}} = {v_{1,x}} - 3 ({\bf q}{\mathcal M}_0{\bf
q}^*)_{x}.
\end{array}
% \right.
\end{equation}
This system admits the real version (${\mathcal
M}^{\top}_0={\mathcal M}^*_0$, ${\bf q}^*={\bf q}^{\top}$)
\begin{equation}\label{cmKDV}
%\left\{
\begin{array}{c}
\alpha_{3}{\bf q}_{t_3}  =  {\bf q}_{xxx} + v_{1}{\bf q}_{x} +
\frac12 v_{1,x}{\bf q},\,\,\,\, \beta_{1}  {v_{1,\tau_1}} =
{v_{1,x}} - 3 ({\bf q}{\mathcal M}_0{\bf q}^{\top})_{x}.
\end{array}
% \right.
\end{equation}

\item
$k=2$, $n=2$. (\ref{spa1}) takes the form
\begin{equation}\label{L2M2y}
\begin{array}{l}
\left[ L_2, M_2\right] =0, \quad
 L_2=\beta_{2} \partial_{\tau_{2}} - {D}^{2} - u_{0}
   - {\bf q}{\mathcal M}_0 {D}^{-1} {\bf r}^{\top},\,\,\\
 M_2=\alpha_{2} \partial_{t_{2}} - {D}^{2} - u_{0}.
 \end{array}
\end{equation}
Under the reduction $\alpha_2,\beta_2\in i{\mathbb{R}}$, ${\bf
r}=\bar{\bf q}$, ${\mathcal M}_0=-{\mathcal M}_0^*$,
$u_0=\bar{u}_0:=u$, the operators $L_2$ and $M_2$ in (\ref{L2M2y})
become Hermitian and (\ref{L2M2y}) takes the form
\begin{equation}\label{Yajima-Oikawa2+1}
\begin{array}{l}
\alpha_{2} {\bf q}_{t_2} =  {\bf q}_{xx} + u{\bf q},\qquad
\alpha_{2} u_{t_2} = \beta_{2}u_{\tau_{2}} + 2({\bf q}{\mathcal
M}_0{\bf q}^*)_{x},
\end{array}
\end{equation}
which is a (2+1)-dimensional vector generalization of the Yajima-Oikawa system.

\item $k=2$, $n=3$. Now (\ref{spa1}) becomes
\begin{equation}\label{398}
\begin{array}{l}
L_2=\beta_2\partial_{\tau_2}- D^2-2u_0-{\bf q}{\mathcal
M}_0{D}^{-1}{\bf r}^{\top}, \,\,
\\
M_3=\alpha_3\partial_{t_3}-D^3-3u_0D-\frac32
\left(u_{0,x}+\beta_2D^{-1}\{u_{0,\tau_2}\}+{\bf q}{\mathcal
M}_0{\bf r}^{\top}\right). \,\,\,
\end{array}
\end{equation}
With the additional reduction $\beta_2\in i{\mathbb{R}}$,
$\alpha_3\in{\mathbb{R}}$, $u_0=\bar{u}_0:=u$, ${\mathcal
M}_0=-{\mathcal M}^*_0$ and ${\bf r}=\bar{\bf q}$, this is
equivalent to the following generalization of the Melnikov system
\cite{PHD},
\begin{eqnarray}
\label{3100}
%\left.
%\begin{array}{l}
&& \alpha_3{\bf q}_{t_3}=  {\bf q}_{xxx} + 3 u{\bf q}_x
 +\frac32\left(u_x+\beta_2D^{-1}\{u_{\tau_2}\}+{\bf q}{\mathcal M}_0{\bf q}^*\right){\bf q}, \nonumber   \\
%\left[
&&
\!\Bigl[\!\Bigr.\alpha_3u_{t_3}\!\!-\!\!\frac14u_{xxx}\!-\!\!3uu_x\!\!
  +\!\!\frac34  \!\left({\bf q}{\mathcal M}_0{\bf q}_x^*\!-\!{\bf q}_x{\mathcal M}_0{\bf
  q}^*\right)_x\!\!
% +
% \right.
%\\\quad\quad
\!-\! \frac34 \beta_2\!\left(\!{\bf q}{\mathcal M}_0{\bf
q}^*\right)_{\tau_2}\!\!\Bigl.\Bigr]_{x}\!
   \!=\!\frac34\beta_2^2 u_{{\tau}_2{\tau}_2}\!. \qquad
%\end{array}
%\right.
\end{eqnarray}
\end{enumerate}
\begin{remark}
(\ref{DeS}) and (\ref{Yajima-Oikawa2+1}) are related by a linear change of the independent
variables.
\end{remark}

Thus, for $k=1$ we have the DS-III hierarchy (its first
members are DS-III ($k=1,n=2$) and a special (2+1)-dimensional
extension of mKdV ($k=1,n=3$), see (\ref{DeS}) and (\ref{mKDV}).
For $k=2$, $k=3$, we have (2+1)-dimensional generalizations
of the Yajima-Oikawa (in particular, it contains (\ref{Yajima-Oikawa2+1}) and (\ref{3100}))
and the Melnikov hierarchy \cite{M4}, respectively.

\section{Lax representations for matrix generalizations of Davey-Stewartson systems}
In this section we introduce an essential extension of the Lax
pair (\ref{L}) for the (2+1)-dimensional extended KP hierarchy to
the case of two integro-differential operators. It leads to DS-I
and DS-II systems. It is known that the Davey-Stewartson (DS-I) system
is connected with the non-stationary Dirac operator (see
\cite{Nizhnik80}). By using a matrix differential representation for the
DS system, involving the non-stationary Dirac operator, and the representation
for DS-III in the (2+1)-dimensional k-cKP hierarchy, we obtain the following Lax
integro-differential Lax pair,
\begin{equation}\label{Ll1M2} L_1=\partial_y-{\bf q}{\mathcal
M}_0D^{-1}{\bf r}^{\top},$$$$
M_2=\alpha_2\partial_{t_2}-c_1D^2-c_2\partial^2_y+2c_1S_1+2c_2{\bf
q}{\mathcal M}_0D^{-1}{\bf r}^{\top}_y+2c_2{\bf q}{\mathcal
M}_0D^{-1}{\bf r}^{\top}\partial_y,
\end{equation}
where $\alpha_2,c_1,c_2\in{\mathbb{C}}$. ${\bf q}={\bf q}(x,y,t_2)$,
${\bf r}={\bf r}(x,y,t_2)$ and $S_1=S_1(x,y,t_2)$ are matrix
functions with dimensions $N\times M$ and $N\times N$, respectively.
${\mathcal M}_0$ is a constant $M\times M$ matrix. The Lax equation
$[L_1,M_2]=0$ is equivalent to the following system,
\begin{equation}\label{DSnr}
%\left\{
\begin{array}{c}
 \alpha_2{\bf q}_{t_2}=c_1{\bf q}_{xx}+c_2{\bf q}_{yy}-2c_1S_1{\bf q}-2c_2{\bf
 q}{\mathcal M}_0S_2,\\
 -\alpha_2{\bf r}^{\top}_{t_2}=c_1{\bf r}^{\top}_{xx}+c_2{\bf r}^{\top}_{yy}-2c_1{\bf
 r}^{\top}S_1-2c_2S_2{\mathcal M}_0{\bf r}^{\top},\\
 S_{1y}=({\bf q}{\mathcal M}_0{\bf r}^{\top})_x,\,\,S_{2x}=({\bf r}^{\top}{\bf q})_y.
\end{array}%\right.
\end{equation}
After the reduction $c_1,c_2\in\mathbb{R}$, $\alpha_2\in
i{\mathbb{R}}$; ${\bf r}^{\top}={\bf q}^*$, ${\mathcal
M}_0={\mathcal M}_0^*$, the operators $L_1$  and $M_2$ are
skew-Hermitian and Hermitian, respectively, and (\ref{DSnr}) takes
the form
\begin{equation}\label{DS}
%\left\{
\begin{array}{c}
 \alpha_2{\bf q}_{t_2}=c_1{\bf q}_{xx}+c_2{\bf q}_{yy}-2c_1S_1{\bf q}-2c_2{\bf
 q}{\mathcal M}_0S_2,\\
 S_{1y}=({\bf q}{\mathcal M}_0{\bf q}^*)_x,\,\,S_{2x}=({\bf q}^*{\bf q})_y.
\end{array}
%\right.
\end{equation}
This has the following two interesting subcases:
\renewcommand{\theenumi}{\arabic{enumi}}
\begin{enumerate}
\item
$c_2=0$. Then we have
\begin{equation}\label{DSc2=0}
%\left\{
\begin{array}{c}
 \alpha_2{\bf q}_{t_2}=c_1{\bf q}_{xx}-2c_1S_1{\bf q},\,\,\,
 S_{1y}=({\bf q}{\mathcal M}_0{\bf q}^*)_x.
\end{array}
%\right.
\end{equation}

\item $c_1=0$. Then (\ref{DS}) takes the form
\begin{equation}\label{DSc1=0}
%\left\{
\begin{array}{c}
 \alpha_2{\bf q}_{t_2}=c_2{\bf q}_{yy}-2c_2{\bf
 q}{\mathcal M}_0S_2,\,\,\,\
S_{2x}=({\bf q}^*{\bf q})_y.
\end{array}
%\right.
\end{equation}
\end{enumerate}
The systems (\ref{DSc2=0}) and (\ref{DSc1=0}) are two different
matrix generalizations of the Davey-Stewartson equation (DS-III)
\cite{Zakharov,6SSS,fokas}. In the vector case ($N=1$),
(\ref{DSc2=0}) has been obtained in \cite{OC} as a member of the
DS-III hierarchy.

\begin{remark}
If $N=1$, the change of variables
$\tilde{x} =x+y$, $\tau_1 =\beta_1y$, $\tilde{\bf q}(\tilde{x},\tau_1)={\bf q}(x,y)$,
$v_0(\tilde{x},\tau_1) =-2S_1(x,y)$  maps (\ref{DSc2=0})
%in new variables $\tilde{x}$, $\tau_1$, $t_2$ and $v_0$, $\tilde{\bf q}$ have the similar
to the DS-III equation (\ref{DeS}) (which we obtained from the (2+1)-dimensional k-cKP hierarchy).
\end{remark}

Let us consider (\ref{DS}) in the case where $u:={\bf q}$ and
${\mathcal M}_0:=\mu$ are scalars. Then (\ref{DS}) becomes
\begin{equation}\label{DSuscal}
%\left\{
 \alpha_2u_{t_2}=c_1u_{xx}+c_2u_{yy}-2c_1S_1u-2\mu c_2 S_2 u,\,\,
 S_{1y}=\mu(|u|^2)_x,\,\,S_{2x}=(|u|^2)_y.
%\right.
\end{equation}
Setting $c_1=c_2=1$ and $\mu=1$, as a consequence of (\ref{DSuscal}) we obtain
\begin{equation}\label{DS2}
%\left\{
\begin{array}{c}
 \alpha_2u_{t_2}=u_{xx}+u_{yy}-2Su,\,\,\,
 S_{xy}=(|u|^2)_{xx}+(|u|^2)_{yy},
\end{array}
%\right.
\end{equation}
where $S=S_1+S_2$. This is the well-known Davey-Stewartson
system (DS-I) and (\ref{DS}) is therefore a matrix generalization.

%Set $c_1=-c_2=1$, $\mu=1$ in system (\ref{DSuscal}). By making the
%change $\tilde{x}=x-y$, $\tilde{y}=x+y$   and putting
%$\tilde{S}(\tilde{x},\tilde{y},t_2):=S_1(x,y,t_2)-S_2(x,y,t_2)$,
%$\tilde{u}(\tilde{x},\tilde{y},t_2):=u(x,y,t_2)$; $\tilde{x}:=x,$
%$\tilde{y}:=y$ we obtain the system:

%\begin{equation}\label{DSuscal2}
%%\left\{
%\begin{array}{c}
% \alpha_2\tilde{u}_{t_2}=4\tilde{u}_{xy}-2\tilde{u}\tilde{S},\\
% \tilde{S}_{yy}-\tilde{S}_{xx}=4|\tilde{u}|^2_{xy}.
%\end{array}%\right.
%\end{equation}

%Systems (\ref{DS2}) and (\ref{DSuscal2}) are different realizations
%of the first Davey-Stewartson system (DS-I).

Now we present integro-differential Lax pairs for
generalizations of the second Davey-Stewartson equation (DS-II).
By replacing the real variables $x$ and $y$ in the operators $L_1$, $M_2$ in
(\ref{Ll1M2}) by a complex variable $z=x + i y$ and its conjugate $\bar{z}$, respectively,
we obtain the following pair,
\begin{equation}\nonumber
 L_1=\partial_{\bar{z}}-{\bf q}{\mathcal M}_0D_z^{-1}{\bf r}^{\top},
 $$$$
 M_2=\alpha_2\partial_{t_2}-c_1D^2_{zz}-c_2\partial^2_{\bar{z}\bar{z}}+2c_1S_1+2c_2{\bf
q}{\mathcal M}_0D_z^{-1}{\bf r}^{\top}_{\bar{z}}+2c_2{\bf
q}{\mathcal M}_0D_z^{-1}{\bf r}^{\top}\partial_{\bar{z}},
\end{equation}
where ${\bf q}={\bf q}(z,\bar{z},t_2)$, ${\bf r}={\bf
r}(z,\bar{z},t_2)$ and $S_1=S_1(z,\bar{z},t_2)$ are matrix functions
with dimensions $N\times M$ and $N\times N$, respectively.
${\mathcal M}_0$ is a constant $M\times M$ matrix. The Lax equation
$[L_1,M_2]=0$ is equivalent to the system
\begin{equation}\label{DSnrz}
%\left\{
\begin{array}{c}
 \alpha_2{\bf q}_{t_2}=c_1{\bf q}_{zz}+c_2{\bf q}_{\bar{z}\bar{z}}-2c_1S_1{\bf q}-2c_2{\bf
 q}{\mathcal M}_0S_2,\\
 -\alpha_2{\bf r}^{\top}_{t_2}=c_1{\bf r}^{\top}_{zz}+c_2{\bf r}^{\top}_{\bar{z}\bar{z}}-2c_1{\bf
 r}^{\top}S_1-2c_2S_2{\mathcal M}_0{\bf r}^{\top},\\
 S_{1\bar{z}}=({\bf q}{\mathcal M}_0{\bf r}^{\top})_z,\,\,S_{2z}=({\bf r}^{\top}{\bf q})_{\bar{z}}.
\end{array}
%\right.
\end{equation}

In terms of the real variables $t_2$, $x$ and $y$, after setting
$c_1=c_2=1$, the system (\ref{DSnrz}) becomes
%a). ��������� ������ � ������� (\ref{DSnr}): $x=z_1+iz_2$,
%$y=z_1-iz_2$,
%$q(x,y,t)=q(z_1+iz_2,z_1-iz_2,t)={\tilde{q}}(z_1,z_2,t)$,
%$S_j(x,y,t)=S_j(z_1+iz_2,z_1-iz_2,t)={\tilde{S}}_j(z_1,z_2,t)$.
%�������� $c_1=c_2=1$, ���������:
\begin{equation}\label{DSnrzz}
%\left\{
\begin{array}{c}
 2\alpha_2{ {\bf q}}_{t_2}={ {\bf q}}_{xx}-{ {\bf q}}_{yy}-4{ {S}}_1{ {\bf q}}-4{ {\bf
 q}}{\mathcal M}_0{ {S}}_2,\\
 -2\alpha_2 {{\bf r}}^{\top}_{t_2}= {{\bf r}}^{\top}_{xx}- {{\bf r}}^{\top}_{yy}-4 {{\bf
 r}}^{\top} {S}_1-4 {S}_2{\mathcal M}_0 {{\bf r}}^{\top},\\
  {S}_{1x}+i {S}_{1y}=({ {\bf q}}{\mathcal M}_0{ {\bf r}}^{\top})_{x}-i({ {\bf q}}{\mathcal M}_0{ {\bf r}}^{\top})_{y},\,\,
  {S}_{2x}-i {S}_{2y}=({ {\bf r}^{\top} {\bf q}})_{x}+i({ {\bf r}^{\top}{\bf q}})_{y}.
\end{array}%\right.
\end{equation}

If $N=M$ (so that ${\bf q}$ and ${\bf r}$ are square matrices), and
$\alpha_2\in i{\mathbb{R}} $, (\ref{DSnrzz}) admits the reduction ${
{\mathcal M}_0{\bf r}^{\top}}={ \bar{\bf q}}$,
${S}_1={\bar{{S}_2}}$, and then takes the following form,
\begin{equation}\label{DSnrzzz}
%\left\{
\begin{array}{c}
 2\alpha_2{ {\bf q}}_{t_2}={ {\bf q}}_{xx}-{ {\bf q}}_{yy}-4{ {S}}_1{ {\bf q}}-4{ {\bf
 q}}{\bar{S}}_1,\\
  {S}_{1x}+i {S}_{1y}=({ {\bf q}}{\bar{\bf q}})_{x}-i({ {\bf q}{\bar{\bf
 q}}})_{y}.
% \tilde{S}_{2z_1}-i\tilde{S}_{2z_2}=({\tilde{\bf r}\tilde{\bf q}})_{z_1}+i({\tilde{\bf r}\tilde{\bf q}})_{z_2}.
\end{array}%\right.
\end{equation}
In the scalar case ($N=M=1$), writing $u={\bf q}$, we obtain the following
consequence of (\ref{DSnrzzz}),
\begin{equation}\label{rs}
%\left\{
\begin{array}{c}
 2\alpha_2{u}_{t_2}={u}_{xx}-{ u}_{yy}-8{\hat{S}}{u}, \quad
  {\hat{S}}_{xx}+{\hat{S}}_{yy}=|u|^2_{xx}-|u|^2_{yy},
% \tilde{S}_{2z_1}-i\tilde{S}_{2z_2}=({\tilde{\bf r}\tilde{\bf q}})_{z_1}+i({\tilde{\bf r}\tilde{\bf q}})_{z_2}.
\end{array}%\right.
\end{equation}
where ${\hat{S}}={\rm Re}({S}_1)$.
This is the second Davey-Stewartson system (DS-II).

\section{Integro-differential Lax representations for a (2+1)-dimensional matrix generalization
of the mKdV equation}

In this section we generalize the Lax representation (\ref{L1M3e}) for the
(2+1)-dimensional mKdV equation to the case of two
integro-differential operators. More precisely, we consider an
extension of the operator $M_3$ in (\ref{L1M3e}). As in the case of the
operator $M_2$ in (\ref{Ll1M2}), by using a differential representation
for the (2+1)-dimensional mKdV equation that involves the non-stationary Dirac operator
\cite{Nizhnik80}, we obtain the following integro-differential Lax pair,
\begin{eqnarray}\label{L1M3}
%\begin{array}{l}
 L_1 &=& \partial_y-{\bf q}{\mathcal M}_0D^{-1}{\bf r}^{\top}, \nonumber \\
 M_3 &=& \alpha_3\partial_{t_3}+c_1D^3-c_2\partial_y^3- 3c_1v_1D-
3c_1v_3+3c_2{\bf q}_y{\mathcal M}_0D^{-1}{\bf r}^{\top}_y \nonumber \\
     && -3c_2{\bf q}{\mathcal M}_0\partial_yD^{-1}{\bf r}^{\top}{\bf q}{\mathcal M}_0D^{-1}{\bf r}^{\top}
  + 3c_2{\bf q}{\mathcal M}_0D^{-1}\{{\bf r}^{\top}{\bf q}\}_y{\mathcal M}_0D^{-1}{\bf r}^{\top} \nonumber \\
 && +3c_2\partial_y{\bf q}{\mathcal M}_0D^{-1}{\bf r}^{\top}\partial_y,
%\end{array}
\end{eqnarray}
%\begin{equation}
%M_3=\alpha_3\partial_{t_3}+c_1D^3-c_2\partial_y^3- 3c_1v_1D-
%3c_1v_3+3c_2{\bf q}_y{\cal M}_0D^{-1}{\bf r}^{\top}_y+3c_2{\bf
%}_y{\cal M}_0D^{-1}{\bf r}^{\top}{\bf q}{\cal M}_0D^{-1}{\bf
%r}^{\top}+$$$$+ 3c_2{\bf q}v_2{\cal M}_0D^{-1}{\bf
%r}^{\top}+3c_2\partial_y{\bf q}{\cal M}_0D^{-1}{\bf
%r}^{\top}\partial_y-3c_2\partial_y{\bf q}{\cal M}_0D^{-1}{\bf
%r}^{\top}{\bf q}{\cal M}_0D^{-1}{\bf r}^{\top},
%\end{equation}
where ${\bf q}={\bf q}(x,y,t_3)$, ${\bf r}={\bf r}(x,y,t_3)$ and
$v_1=v_1(x,y,t_3),\,\,v_3=v_3(x,y,t_3)$ are $N\times M$, respectively $N\times N$, matrix functions.
As in the case of the pair $L_1$, $M_2$ in (\ref{Ll1M2}), a change of variables shows that in
the case $c_2=0$ and $N=1$ the Lax pairs (\ref{L1M3}) and (\ref{L1M3e}) are equivalent.
The Lax equation $[L_1,M_3]=0$ results in the system
\begin{eqnarray}\label{DSL1M3}
&& \alpha_3{\bf q}_{t_3}+c_1{\bf q}_{xxx}-c_2{\bf q}_{yyy}-
3c_1v_1{\bf q}_x+ 3c_2{\bf q}_y{\mathcal M}_0v_2+ 3c_2{\bf
q}{\mathcal M}_0v_{2y}\nonumber \\ &&-3c_1v_3{\bf q}-3c_2{\bf
q}{\mathcal M}_0v_4  -3c_2{\bf q}D^{-1}\left\{{\mathcal M}_0{\bf
r}^{\top}{\bf q}{\mathcal M}_0v_2
  -{\mathcal M}_0v_2{\mathcal M}_0{\bf r}^{\top}{\bf q}\right\}=0, \nonumber \\
&& \alpha_3{\bf r}^{\top}_{t_3}+c_1{\bf r}^{\top}_{xxx}-c_2{\bf
r}^{\top}_{yyy}- 3c_1{\bf r}^{\top}_xv_1- 3c_1{\bf r}^{\top}v_{1x}+
3c_2v_2{\mathcal M}_0{\bf r}^{\top}_y\nonumber\\&&+ 3c_1{\bf
r}^{\top}v_3 +3c_2v_4{\mathcal M}_0{\bf r}^{\top}-
3c_2D^{-1}\left\{v_2{\mathcal M}_0{\bf r}^{\top}{\bf q}-{\bf
r}^{\top}{\bf q}{\mathcal
M}_0v_2\right\}{\mathcal M}_0{\bf r}^{\top}=0, \nonumber \\
&& v_{1y}=({\bf q}{\mathcal M}_0{\bf r}^{\top})_x,\,
v_{2x}=({\bf{r}}^{\top}{\bf q})_y,\,\nonumber\\&& v_{3y}=({\bf
q}_x{\mathcal M}_0{\bf r}^{\top})_x+ [{\bf q}{\mathcal M}_0{\bf
r}^{\top},v_1],\,v_{4x}=({\bf r}^{\top}_y{\bf q})_y.
\end{eqnarray}
Let us consider some reductions of this system:
\begin{enumerate}
%1). ��� $c_2=0$ ������� (\ref{DSL1M3}) ������ �����\cite{ws}��:
%\begin{equation}
%\alpha_3{\bf q}_{t_3}+c_1{\bf q}_{xxx}- 3c_1v_1{\bf q}_x-
%3c_1v_3{\bf q}=0,
%$$$$
%\alpha_3{\bf r}^{\top}_{t_3}+c_1{\bf r}^{\top}_{xxx}- 3c_1{\bf
%r}^{\top}_xv_1- 3c_1{\bf r}^{\top}v_{1x}+ 3c_1{\bf r}^{\top}v_3=0,
%$$$$
% v_{1y}=({\bf
%q}{\cal M}_0{\bf r}^{\top})_x,\, v_{3y}=({\bf q}_x{\cal M}_0{\bf
%r}^{\top})_x+[{\bf q}{\cal M}_0{\bf r}^{\top},v_1].
%\end{equation}
%2). ��� $c_1=0$ ������� (\ref{DSL1M3}) ������ �������:%%
%\begin{equation}\nonumber
%alpha_3{\bf q}_{t_3}-c_2{\bf q}_{yyy}+3c_2{\bf q}_y{\cal M}_0v_2+
%3c_2{\bf q}{\cal M}_0v_{2y}-3c_2{\bf q}{\cal M}_0v_4-$$$$-3c_2{\bf
%q}D^{-1}\left\{{\cal M}_0{\bf r}^{\top}{\bf q}{\cal M}_0v_2-{\cal
%M}_0v_2{\cal M}_0{\bf r}^{\top}{\bf q}\right\}=0,
%$$$$
%\alpha_3{\bf r}^{\top}_{t_3}-c_2{\bf r}^{\top}_{yyy}+ 3c_2v_2{\cal
%M}_0{\bf r}^{\top}_y+3c_2v_4{\cal M}_0{\bf r}^{\top}-
%3c_2D^{-1}\left\{v_2{\cal M}_0{\bf r}^{\top}{\bf q}-{\bf
%r}^{\top}{\bf q}{\cal M}_0v_2\right\}{\cal M}_0{\bf r}^{\top}=0,$$$$
% v_{2x}=({\bf{r}}^{\top}{\bf q})_y,\,\,v_{4x}=({\bf
%r}^{\top}_y{\bf q})_y
%\end{equation}
\item $\alpha_3,c_1,c_2\in{\mathbb{R}}$, ${\bf r}^{\top}={\bf q}^*$, ${\mathcal M}_0={\mathcal M}^*_0$.
The operators $L_1$ and $M_3$ are then skew-Hermitian and (\ref{DSL1M3}) takes the form
\begin{eqnarray}
&& \alpha_3{\bf q}_{t_3}+c_1{\bf q}_{xxx}-c_2{\bf q}_{yyy}-
3c_1v_1{\bf q}_x+ 3c_2{\bf q}_y{\mathcal M}_0v_2+ 3c_2{\bf
q}{\mathcal M}_0v_{2y}\nonumber\\&& -3c_1v_3{\bf q}-3c_2{\bf
q}{\mathcal M}_0v_4- 3c_2{\bf q}D^{-1}\left\{{\mathcal M}_0{\bf
q}^{*}{\bf q}{\mathcal M}_0v_2-{\mathcal
M}_0v_2{\mathcal M}_0{\bf q}^{*}{\bf q}\right\}=0, \nonumber \\
&&  v_{1y}=({\bf q}{\mathcal M}_0{\bf q}^*)_x,\, v_{2x}=({\bf
q}^*{\bf{q}})_y,\,\nonumber\\&& v_{3y}=({\bf q}_x{\mathcal M}_0{\bf
q}^*)_x+ [{\bf q}{\mathcal M}_0{\bf q}^*,v_1],\,v_{4x}=({\bf
q}^*_y{\bf q})_y. \label{DSL1M3r}
\end{eqnarray}
In the scalar case ($N=M=1$), setting ${\mathbb{R}}\ni\mu:={\mathcal
M}_0$, $q(x,y,t_3):={\bf q}(x,y,t_3)$, (\ref{DSL1M3r}) reads
\begin{eqnarray}\nonumber
&& \alpha_3{q}_{t_3}+c_1{q}_{xxx}-c_2{q}_{yyy}-
3c_1\mu{q}_x\int|q|^2_xdy+\nonumber\\ && 3c_2\mu{q}_y\int|q|^2_ydx+
3c_2\mu{q}\int({\bar{q}}q_y)_ydx-3c_1\mu{q}\int({q}_xq)_xdy=0.
\end{eqnarray}
%3a). ���� �������� $c_2=0$, �� ������� (\ref{DSL1M3r}) ������
%�������
%\begin{equation}\nonumber
%\alpha_3{\bf q}_{t_3}+c_1{\bf q}_{xxx}- 3c_1v_1{\bf q}_x-
%3c_1v_3{\bf q}=0,
%$$$$
%  v_{1y}=({\bf
%q}{\cal M}_0{\bf q}^*)_x,\, v_{3y}=({\bf q}_x{\cal M}_0{\bf q}^*)_x+
%[{\bf q}{\cal M}_0{\bf q}^*,v_1].
%\end{equation}
%3b). � ������� $c_1=0$ ������� (\ref{DSL1M3r}) ���� �����:%
%
%\begin{equation}\nonumber
%\alpha_3{\bf q}_{t_3}-c_2{\bf q}_{yyy}+ 3c_2{\bf q}_y{\cal M}_0v_2+
%3c_2{\bf q}{\cal M}_0v_{2y}-3c_2{\bf q}{\cal M}_0v_4-$$$$-3c_2{\bf
%q}D^{-1}\left\{{\cal M}_0{\bf q}^{*}{\bf q}{\cal M}_0v_2-{\cal
%M}_0v_2{\cal M}_0{\bf q}^{*}{\bf q}\right\}=0,
%$$$$
% v_{2x}=({\bf q}^*{\bf{q}})_y,\,v_{4x}=({\bf q}^*_y{\bf
%q})_y.
%\end{equation}
In the real case ${\bf q}=\bar{{\bf q}}$, (\ref{DSL1M3r}) becomes
\begin{eqnarray}\label{DSL1M3rd}
&& \alpha_3{\bf q}_{t_3}+c_1{\bf q}_{xxx}-c_2{\bf q}_{yyy}-
3c_1v_1{\bf q}_x+ 3c_2{\bf q}_y{\mathcal M}_0v_2+ 3c_2{\bf
q}{\mathcal M}_0v_{2y}\nonumber\\
&&-3c_1v_3{\bf q}-3c_2{\bf q}{\mathcal M}_0v_4 - 3c_2{\bf
q}D^{-1}\left\{{\mathcal M}_0{\bf q}^{{\top}}{\bf q}{\mathcal
M}_0v_2-{\mathcal
M}_0v_2{\mathcal M}_0{\bf q}^{{\top}}{\bf q}\right\}=0, \nonumber \\
&&   v_{1y}=({\bf q}{\mathcal M}_0{\bf q}^{\top})_x,\, v_{2x}=({\bf
q}^{\top}{\bf{q}})_y,\, v_{3y}=({\bf q}_x{\mathcal M}_0{\bf
q}^{\top})_x+[{\bf q}{\mathcal M}_0{\bf
q}^{\top},v_1],\,\nonumber\\&&v_{4x}=({\bf q}^{\top}_y{\bf q})_y.
\qquad
\end{eqnarray}
In the scalar case ($N=M=1$), writing ${\mathcal M}_0=\mu$ and
$q=q(x,y,t_3)$=${\bf q}(x,y,t_3)$,  after setting $y=x$ and
$c_1-c_2=1$, (\ref{DSL1M3rd}) has the form
%\begin{equation}\label{DSL1M3rddd}
%\alpha_3{ q}_{t_3}+c_1{ q}_{xxx}-c_2{ q}_{yyy}- 3c_1v_1{q}_x+ 3\mu
%c_2{ q}_yv_2+ 3\mu c_2{ q}v_{2y}- 3c_1v_3{ q}-3\mu c_2{ q}v_4=0,
%$$$$
%  v_{1y}=2\mu qq_x,\, v_{2x}=2qq_y,\,
%v_{3y}=\mu(q_xq)_x,\,v_{4x}=(q_yq)_y.
%\end{equation}
%� ���������� ������� $N=M=1$, $\mu:={\cal M}_0$ �������
%(\ref{DSL1M3rd}) ������ �������:
%\begin{equation}\label{DSL1M3r2+1}
%\alpha_3{q}_{t_3}+c_1{q}_{xxx}-c_2{q}_{yyy}- 3c_1\mu(
%_x+\frac12q\partial_x)\int q^2_xdy+
%c_2\mu(q_y+\frac12q\partial_y)\int q^2_ydx=0.
%\end{equation}
%��� $y=x$, $c_1-c_2=1$ �������� (\ref{DSL1M3r2+1}) ������ ������
%�������:
\begin{equation}\label{DSL1M3rdddd}
\alpha_3{ q}_{t_3}+{q}_{xxx}- 6\mu q^2q_x=0,
\end{equation}
which is the mKdV equation. The systems (\ref{DSL1M3r}) and (\ref{DSL1M3rd}) are therefore,
respectively, complex and real, spatially two-dimensional matrix generalizations of it.

\item ${\mathcal M}_0{\bf r}^{\top}=\nu$ with a constant matrix $\nu$.
In terms of $u := {\bf q}\nu$, (\ref{DSL1M3}) takes the form
\begin{eqnarray}\label{DSL1M3r2}
&& \alpha_3u_{t_3}+c_1u_{xxx}-c_2u_{yyy}-3c_1D\left\{\left(\int u_x
dy\right)u\right\}\nonumber \\&&+3c_2\partial_y\left\{u\left(\int
u_y
dx\right)\right\}  - 3c_1\left(\int[u,v_1]dy\right)u-3c_2u\left(\int[u,v_2]dx\right)=0, \nonumber \\
&&  \nu\left(c_1\int[u,v_1]dy-c_2\int[v_2,u]dx\right)=0, \quad
  v_{1y}=u_x,\,\,\, v_{2x}=u_y.
% 3c_1v_{3y}=3c_1{\bf
%q}_{xx}\nu+ 3c_1[{\bf q}\nu,v_1],\,v_{4x}=0.
\end{eqnarray}
%������� ������� ���� ���� ���������� � ������ �������:
%\begin{equation}\label{NDS}
%\nu\left(\alpha_3{\bf q}_{t_3}+c_1{\bf q}_{xxx}-c_2{\bf
%Z%q}_{yyy}-3c_1D\left(\int{\bf q}_x dy\right)\nu{\bf
%q}+3c_2\partial_y{\bf q}\nu\left(\int{\bf q}_y
%dx\right)\right)+$$$$+ 3c_2[\nu{\bf q},D^{-1}\{[\nu{\bf
%q},v_2]\}]=0,
%$$$$
%\nu 3c_1\left(\int\{[{\bf q}\nu,v_1]\}dy\right)+
%3c_2D^{-1}\{[v_2,{\nu}{\bf q}]\}{\nu}=0,$$$$
%  v_{1y}={\bf
%q}_x\nu,\,v_{2x}=\nu{\bf{q}}_y.
%\end{equation}
%4a). ���� $c_2=0$, �� ������� (\ref{DSL1M3r2}) ����������� ���:
%\begin{equation}\label{DSL1Ms2r22}
%\alpha_3u_{t_3}+c_1u_{xxx}-3c_1D\left\{\left(\int u_x
%dy\right)u\right\}-3c_1\left(\int[u,v_1]dy\right)u=0,
%$$$$
% \nu\int[u,v_1]dy=0,
%  v_{1y}=u_x.% 3c_1v_{3y}=3c_1{\bf
%%q}_{xx}\nu+ 3c_1[{\bf q}\nu,v_1],\,v_{4x}=0.
%\end{equation}
%4b). � ������� $c_1=0$ ������� (\ref{DSL1M3r2}) ������ ������
%�������:
%\begin{equation}\label{DSL1M3r23}
%\alpha_3u_{t_3}-c_2u_{yyy}+3c_2\partial_y\left\{u\left(\int u_y
%dx\right)\right\}+3c_2uD^{-1}\{[u,v_2]\}=0,
%$$$$
% \nu D^{-1}\{[v_2,u]\}=0,\, v_{2x}=u_y.% 3c_1v_{3y}=3c_1{\bf
%%q}_{xx}\nu+ 3c_1[{\bf q}\nu,v_1],\,v_{4x}=0.
%\end{equation}%
In the scalar case ($N=1,M=1$), this reduces to
\begin{equation}\label{DSL1M3r2scal}
\alpha_3u_{t_3}+c_1u_{xxx}-c_2u_{yyy}-3c_1D\left\{\left(\int u_x
dy\right)u\right\}+3c_2\partial_y\left\{u\left(\int u_y
dx\right)\right\}=0,
\end{equation}
which is the Nizhnik equation \cite{Nizhnik80}. The system (\ref{DSL1M3r2}) thus generalizes
the latter to the matrix case.
%-��������-��������
%,VesNov,ital}.
\end{enumerate}

\section{Integro-differential Lax representation for a spatially two-dimensional
matrix generalization of the Chen-Lee-Liu equation}

 In this section we apply a gauge transformation to the Lax
pair (\ref{Ll1M2}) in order to obtain an integro-differential Lax pair
for the (2+1)-dimensional Chen-Lee-Liu equation. The resulting Lax pair generalizes
a corresponding Lax pair obtained from the (2+1)-dimensional k-cmKP hierarchy.
Let $f=f(x,y,t_2)$ be an $M\times M$ matrix solution of $L_1\{f\}=0$, where the
operator $L_1$ has the form given in (\ref{Ll1M2}). Consider the following gauge transformation,
\begin{eqnarray}\nonumber
&&f^{-1}\!L_1f=f^{-1}\!\left(\partial_y\!-\!{\bf q}{\mathcal
M}_0D^{-1}{\bf r}^{\top}\!\right)\!f=\!\partial_y+\nonumber\\
&&+f^{-1}{\bf q}{\mathcal M}_0D^{-1}D^{-1}\{{\bf
r}^{\top}f\}D:=\partial_y-\tilde{\bf q}{\mathcal
M}_0D^{-1}\tilde{\bf r}^{\top}D,
\end{eqnarray}
where $\tilde{\bf q}:=f^{-1}{\bf q}$, $\tilde{\bf
r}^{\top}:=-D^{-1}\{{\bf r}^{\top}f\}$. It is also possible to apply a
similar transformation to the operator $M_2$ in (\ref{Ll1M2}). In order
to simplify notation, in the following we write ${\bf q},{\bf r}$ instead of
$\tilde{{\bf q}},\tilde{{\bf r}}$.
As a result of the gauge transformation, we obtain a Lax pair of the form
\begin{equation}\label{DL1}
L_1=\partial_y-{\bf q}{\mathcal M}_0D^{-1}{\bf r}^{\top}D,
\end{equation}
\begin{equation}\label{DM2}
M_2=\alpha_2\partial_{t_2}-c_1D^2-c_2\partial_y^2+2c_1S_1D+2c_2{\bf
q}{\mathcal M}_0D^{-1}\partial_y{\bf r}^{\top}D,
\end{equation}
where ${\bf q}={\bf q}(x,y,t_2)$, ${\bf r}={\bf r}(x,y,t_2)$ and
$S_1=S_1(x,y,t_2)$ are $N\times M$, respectively $N\times N$, matrix
functions. ${\mathcal M}_0$ is a constant $M\times M$ matrix. The
condition $[L_1,M_2]=0$ is then equivalent to the following system,
\begin{equation}\label{sysD2}
%\left\{
\begin{array}{c} \alpha_2{\bf q}_{t_2}-c_1{\bf
q}_{xx}-c_2{\bf q}_{yy}+2c_1S_1{\bf q}_x-2c_2{\bf
q}{\mathcal M}_0S_2+2c_2{\bf q}{\mathcal M}_0({\bf r}^{\top}{\bf q})_y=0,\\
\left(\alpha_2{\bf r}^{\top}_{t_2}+c_1{\bf r}^{\top}_{xx}+c_2{\bf
r}^{\top}_{yy}+2c_1{\bf r}^{\top}_xS_1+2c_2S_2{\mathcal M}_0{\bf
r}^{\top}\right)_x=0,\\
S_{1y}=({\bf q}{\mathcal M}_0{\bf r}^{\top})_x+[{\bf q}{\mathcal
M}_0{\bf r}^{\top},S_1],\,\,S_{2x}=({\bf r}^{\top}_x{\bf q})_y.
\end{array}
%\right.
\end{equation}
Let us consider some of its reductions:
\begin{enumerate}
\item
 The system (\ref{sysD2}) contains three different (2+1)-dimensional matrix generalizations
 of the Chen-Lee-Liu system: a) $c_1=0$,$\quad$ b) $c_2=0$, $\quad$ c) $c_1\neq0$ and
 $c_2\neq0$. \\
In case b), $M_2$ becomes a differential operator. With the restriction $N=1$, this
case appeared in \cite{OC}.

\item
  $\alpha_2\in i{\mathbb{R}}$,
$c_1,c_2\in{\mathbb{R}}$, ${\mathcal M}_0=-{\mathcal M}^*_0$, ${\bf
r}^{\top}={\bf q}^*$, $S_1=S_1^*$. Then $L_1$ and $M_2$, given by
(\ref{DL1}) and (\ref{DM2}), are $D$-skew-Hermitian
($L^*_1=-DL_1D^{-1}$) and $D$-Hermitian ($M^*_2=DM_2D^{-1}$),
respectively. (\ref{sysD2}) has the following form,
\begin{equation}\label{sysD2r1}
%\left\{
\begin{array}{c} \alpha_2{\bf q}_{t_2}-c_1{\bf
q}_{xx}-c_2{\bf q}_{yy}+2c_1S_1{\bf q}_x-2c_2{\bf
q}{\mathcal M}_0S_2+2c_2{\bf q}{\mathcal M}_0({\bf q}^{*}{\bf q})_y=0,\\
S_{1y}=({\bf q}{\mathcal M}_0{\bf q}^{*})_x+[{\bf q}{\mathcal
M}_0{\bf q}^{*},S_1],\,\,S_{2x}=({\bf q}^{*}_x{\bf q})_y.
\end{array}
%\right.
\end{equation}
In the scalar case ($N=M=1$), setting $c_1=1$, $c_2=0$, $y=x$, and
writing $\mu={\mathcal M}_0$, $q=q(x,y,t_2)={\bf q}(x,y,t_2)$,
(\ref{sysD2r1}) reduces to the Chen-Lee-Liu equation \cite{Chen}
\begin{equation}\nonumber
\alpha_2{q}_{t_2}-{q}_{xx}+2\mu|q|^2q_x=0.
\end{equation}

\item ${\mathcal M}_0{\bf r}^{\top}=\nu$ with a constant matrix $\nu$. Then
the operators (\ref{DL1}) and (\ref{DM2}) are purely differential and,
in terms of $u:={\bf q}\nu$, (\ref{sysD2}) reads
%\begin{equation}\label{sysD2r2}
%\left\{\begin{array}{c} \alpha_2{\bf q}_{t_2}-c_1{\bf
%q}_{xx}-c_2{\bf q}_{yy}+2c_1S_1{\bf q}_x+2c_2{\bf q}\nu{\bf q}_y=0,\\
%S_{1y}={\bf q}_x\nu+[{\bf q}\nu,S_1].
%\end{array}\right.
%\end{equation}
\begin{equation}\label{sysD2r2}
%\left\{
\begin{array}{c}
\alpha_2u_{t_2}-c_1u_{xx}-c_2u_{yy}+2c_1S_1u_x+2c_2uu_y=0, \qquad
S_{1y}=u_x+[u,S_1].
\end{array}%\right.
\end{equation}
This is a spatially two-dimensional matrix generalization of the Burgers equation.
A generalization of (\ref{sysD2r2}) to an arbitrary number of
spatial dimensions has been considered in \cite{CM}.
\end{enumerate}
It is possible to apply a similar gauge transformation
% as it was done for Lax pair (\ref{Ll1M2})
to the pair of operators (\ref{L1M3}).
Corresponding Lax representations that lead to matrix versions of spatially
two-dimensional higher Chen-Lee-Liu equations were investigated in \cite{CM}.

\section{Conclusions}

In this paper we considered several members of a (2+1)-dimensional
generalization of the k-cKP hierarchy (\ref{spa1}). Originally, this
substantial generalization of the k-cKP hierarchy had been proposed
in \cite{MSS,6SSS}. For some members of this hierarchy (e.g.
(\ref{Yajima-Oikawa2+1}) and (\ref{3100})), solutions were obtained
via the binary
%%% Darboux ???
transformation dressing method \cite{BS1,PHD}. The (2+1)-dimensional
extension of the k-cKP hierarchy has been rediscovered more recently
in \cite{LZL1}, also see \cite{LZL2,YYQB} for further
investigations. The  authors of \cite{LZL1} and \cite{LZL2}
considered the (2+1)-dimensional k-cKP and (2+1)-dimensional k-cmKP
hierarchies with
%%% Check "the" !
two operators of the Lax pair having different orders of
differentiation with respect to $x$. This excludes, for example,
(2+1)-dimensional vector generalizations of the Yajima-Oikawa
(\ref{Yajima-Oikawa2+1}) and Drinfeld-Sokolov-Wilson equation
\cite{DS,WG,HGR}:
\begin{equation}\label{DSW}
{\bf q}_{t_3}={\bf q}_{xxx}+3u{\bf q}_x+\frac32u_x{\bf q},\,\,
u_{t_3}=u_{\tau_3}+3({\bf q}{\mathcal M}_0{\bf q}^{\top})_x.
\end{equation}
It is obtained from $[L_3,M_3]=0$ with
\begin{equation}
L_3=\partial_{\tau_3}-D^3-3uD-\frac32 u_x-{\bf q}{\mathcal
M}_0D^{-1}{\bf q}^{\top},\quad
M_3=\partial_{t_3}-D^3-3uD-\frac32u_x.
\end{equation}
%Equation (\ref{DSW}) describes the evolution and interaction of two
%types of plane waves on the plane $(x,\tau_3)$.
%%% Reference or further explanation ?
\begin{remark}
Analogously to Remark 2.1, it can be shown that the (2+1)-dimensional
vector generalization of the Drinfeld-Sokolov-Wilson equation
(\ref{DSW}) is equivalent to the real version of the (2+1)-dimensional
multi-component mKdV equation (\ref{cmKDV}) via a linear change of
independent variables.
\end{remark}
%%% Delete the following paragraph ???
%Despite the fact that (2+1)-dimensional vector generalizations of the
%Yajima-Oikawa and the Drinfeld Sokolov-Wilson systems are connected with
%the DS-III equations (\ref{DeS}) and the (2+1)-dimensional generalization (\ref{cmKDV})
%of real mKdV by a linear change of variables (see Remarks
%1.1 and 6.1), they describe physically different processes.
%%% Here you

 The aim of our work was also to generalize Lax
representations for members of the (2+1)-dimensional k-cKP hierarchy
\cite{MSS,6SSS,LZL1,LZL2}, in order to obtain integrable equations
that do not belong to the (2+1)-dimensional k-cKP hierarchy. In
particular, we constructed Lax integro-differential representations
for matrix generalizations of the Davey-Stewartson systems
(DS-I,DS-II,DS-III), matrix generalizations of (2+1)-dimensional
extensions of the mKdV, the Nizhnik \cite{Nizhnik80} and the
Chen-Lee-Liu \cite{Chen} equations. Representations for some of
those systems in the algebra of purely differential operators with
%%% check:
matrix coefficients
%dimension
can be found in \cite{11Athorne}.

One of the problems left for further investigation are the dressing
methods for the corresponding Lax representations. It was shown that
for the Lax representations for the (2+1)-dimensional k-cKP
hierarchy \cite{MSS,6SSS,LZL2}, one can use differential operators
for the dressing. The most interesting systems obtained from the
(2+1)-dimensional k-cKP hierarchy and in Sections 3-5 arise after a
Hermitian conjugation reduction. This imposes nontrivial constraints
on the dressing differential operator. It was shown for the k-cKP
hierarchy \cite{OSCH,BS2} that it is more suitable to use a binary
transformation operator in this situation. The dressing method by
binary transformations for evolution integro-differential operators
that arise in the (2+1)-dimensional k-cKP case has been considered
in \cite{SCD}. We plan to apply this method to the systems in
Sections 3-5 in a forthcoming paper. Another interesting question is
the possibility of generalizations of other representations from the
(2+1)-dimensional k-cKP hierarchy to the case of two
integro-differential operators, e.g. (2+1)-dimensional extensions of
the Yajima-Oikawa and the Melnikov systems.
 %By using
%the method presented in \cite{SCD} for dressing
% of the integro-differential operators we can obtain the exact solutions of the systems with integro-differential Lax representations.
%We will apply this method for the above mentioned systems in the
%next paper.
\section{Acknowledgments}
The authors thank Professor M\"uller-Hoissen for
fruitful discussions and useful advice in preparation of this
paper. O. Chvartatskyi also thanks the organizers of the ISQS-20
Conference for the warm hospitality in Prague.


\begin{thebibliography}{29}


\bibitem{LDA} Dickey L A 2003 {\it Soliton Equations and Hamiltonian Systems}, Advanced
Series in Mathematical Physics, 2nd ed. World Scientific, River
Edge, NJ, Vol. 26.

\bibitem{Zakharov} Zakharov V E and Shabat A B 1974 {\it Funct. Anal. Appl.} {\bf{8}} (3), 226

\bibitem{Zakh-Manak} Novikov S P, Manakov S V, Pitaevskij L P and Zakharov V
E 1984
 {\it Theory of solitons. The inverse scattering methods.}
 (Transl. from the Russian), Contemporary Soviet Mathematics. New York - London: Plenum Publishing Corporation. Consultants Bureau,

\bibitem{solitons} Bullough R K and Caudrey P J (eds.), 1980 {\it Solitons},
Springer-Verlag, Berlin

\bibitem{March} {{Marchenko V A}} 1988 {\it Nonlinear equations and operator
algebras,} Dordrecht, Boston, Lancaster, Tokyo, Reidel

\bibitem{Matveev79} {Matveev V B} 1979 {\it Lett. in Math. Phys.} {\bf 3} 213

\bibitem{Matveev}  {Matveev V B and Salle M A} 1991 {\it Darboux transformations and
solitons}, Berlin Heidelberg, Springer-Verlag.

\bibitem{DJKM1} Date E, Jimbo M, Kashiwara M, and Miwa T 1981 {\it J. Phys. Soc.
Jpn.}
{\bf 50} 3806
\bibitem{DJKM2} Date E, Jimbo M, Kashiwara M, and Miwa T 1982 {\it Publ. Res. Inst.
Math. Sci.} {\bf 18} 1077

\bibitem{SS3} Sato M and Sato Y 1983 {\it North-Holland Math. Stud.} {\bf 81} 259

\bibitem{MM4} Jimbo M and Miwa T 1983 {\it Publ. Res. Inst. Math. Sci.} {\bf 19}
943

\bibitem{Ohta} {Ohta Y, Satsuma J, Takahashi D and Tokihiro T}
1988 {\it Prog. Theor. Phys. Suppl.} {\bf 94} 210

\bibitem{M1} Melnikov V K 1983 {\it Lett. Math. Phys.} {\bf 7} 129

\bibitem{Mf} Melnikov V K 1985 {\it Preprint JINR P2-85-958}, Dubna (in
Russian)

\bibitem{M4} Melnikov V K 1987 {\it Preprint JINR P2-87-136}, Dubna (in
Russian)

\bibitem{M2} Melnikov V K 1987 {\it Commun. Math. Phys.} {\bf 112} 639

\bibitem{M3} Melnikov V K 1988 {\it Phys. Lett. A} {\bf 128} 488






\bibitem {SS} Sidorenko J and Strampp W 1991 {\it Inverse Problems} {\bf 7} L37

\bibitem{KSS} Konopelchenko B G, Sidorenko J and Strampp W 1991 {\it Phys.
Lett. A} {\bf 157} 17

\bibitem{Chenga1} Cheng Y and Li Y S 1991 {\it Phys. Lett. A} {\bf 157} 22

\bibitem{CY} {Cheng Y} 1992  {\it J.~Math. Phys.} {\bf33} 3774


\bibitem{Chenga2} Cheng Y and Li Y S 1992 {\it J. Phys. A.} {\bf 25} 419

\bibitem{SSq} {Sidorenko J and Strampp W} 1993 {\it J. Math. Phys.} {\bf 34} (4) 1429

\bibitem{Oevel93} Oevel W 1993 {\it Physica A} {\bf195} 533

\bibitem{Oevel96} Oevel W and Strampp W 1996 {\it J. Math. Phys.} {\bf37} 6213

\bibitem{Aratyn97} Aratyn H, Nissimov E, and Pacheva S 1997 {\it Int. J. Mod. Phys. A} {\bf 12} 1265




%\bibitem{SSBM} {Samoilenko V.H., Sidorenko Yu.M., Buonanno L., Matarazzo G.}
%2000 Proceedings of Institute of Math. of NAS of Ukraine, 30 (Part
%II) 406

\bibitem{KSO} Kundu A, Strampp W and Oevel W 1995 {\it J. Math. Phys.} {\bf
36} 2972



\bibitem{OC} Oevel W and Carillo S 1998 {\it J. Math. Anal. Appl.} {\bf 217} 161


\bibitem{MSS}{Mytropolsky Yu O, Samoilenko V H and Sidorenko Yu M}
1999 {\it Proceedings of NSA of Ukraine} {\bf{8}} 19


\bibitem{6SSS} { Samoilenko A M, Samoilenko V G and Sidorenko Yu M} 1999  {\it Ukr. Math. Journ.} {\bf  51} (1) 86



\bibitem{BS1} Berkela Yu Yu and Sydorenko Yu M 2002 {\it Mat.
Studii}
{\bf25} (1) 38


\bibitem{PHD} Berkela Yu Yu  "Integration of nonlinear evolution systems with nonlocal constraints" Ph.D. thesis, Ivan Franko National University of
Lviv (in Ukrainian)


\bibitem{LZL1} Liu X J, Zeng Y B and  Lin R 2008 {\it Phys. Lett.} {\bf 372} 3819

\bibitem{LZL2} Liu X J, Lin R, Jin B and Zeng Y B 2009 {\it J. Math. Phys.} {\bf 50} 053506

\bibitem{Nizhnik80} {Nizhnik L P} 1980 {\it Dokl. Akad. Nauk SSSR} {\bf 254}  332


\bibitem{SCD} {Sydorenko Yu and Chvartatskyi O} 2009 {\it Visn. Kyiv. Univ. Ser: mech.-mat.} {\bf22}
32 (in Ukrainian)



%\bibitem{ZS}{{V. E. Zakharov, A. B. Shabat.}} 1974 Funkts. Anal. Prilozh.
%{\bf 8} (3) 43

\bibitem{fokas} {Fokas A S} 1994 {\it Inverse Problems} {\bf 10} L19

 \bibitem{Chen}{Chen H H, Lee Y C, Liu C S} 1979  {\it Physica Scr.} {\bf
 20} 490

\bibitem{CM}{Sydorenko Yu and Chvartatskyi O} 2012 {\it Carpathian Mathematical publications} {\bf 4} (1)
124 (in Ukrainian)

\bibitem{YYQB} {Huang Y,Yao Y Q and Zeng Y B} 2012 {\it Commun. Theor.
Phys.} {\bf57} (4) 515




 \bibitem{DS} Drinfeld V G and Sokolov V V 1981 {\it Sov. Math. Dokl.} {\bf 23} 457

\bibitem{WG} Wilson G 1982 {\it Phys. Lett. A} {\bf 89} 332

\bibitem{HGR}
 Hirota R, Grammaticos B and Ramani A 1986 {\it J. Math. Phys} {\bf 27}
 (6) 1499

\bibitem{11Athorne}
{Athorne C, Fordy A P} 1987 {\it J. Math. Phys.} {\bf 28}  2018


\bibitem{OSCH} {Oevel W and Schief W} 1994 {\it Reviews in
Mathematical Physics} {\bf6} 1301

\bibitem{BS2} Berkela Yu Yu and Sydorenko Yu M 2006 {\it Mat.
Studii}
{\bf25} (1) 38 (in Ukrainian)

%\bibitem{WS} {W. Oevel, W. Strampp} 1996 J. Math. Phys. {\bf 37} 6213


\end{thebibliography}
\end{document}